\documentclass{PoS_oldRef}

\title{Structure, dynamical impact and origin of magnetic fields in nearby galaxies in the SKA era}

\ShortTitle{Magnetic fields in nearby galaxies}

\author{\speaker{Rainer Beck} $^1$ \thanks{On behalf of the SKA Cosmic Magnetism Working Group} ,
{Dominik Bomans}$^2$, {Sergio Colafrancesco}$^3$, {Ralf-J\"urgen Dettmar}$^2$, {Katia Ferri\`ere}$^4$,
{Andrew Fletcher}$^5$, {George Heald}$^6$, {Volker Heesen}$^7$,
{Cathy Horellou}$^8$, {Marita Krause}$^1$, {Yu-Qing Lou}$^9$, {Sui Ann Mao}$^1$, {Rosita Paladino}$^{10}$,
{Eva Schinnerer}$^{11}$, {Dmitry Sokoloff}$^{12}$, {Jeroen Stil}$^{13}$ and {Fatemeh Tabatabaei}$^{11}$
\\
$^1$ Max-Planck-Institut f\"ur Radioastronomie, Auf dem H\"ugel 69, 53121 Bonn, Germany\\
$^2$ Astron. Inst. der Ruhr-Universit\"at Bochum, Universit\"atsstr. 150, 44780 Bochum, Germany\\
$^3$ School of Physics, University of the Witwatersrand, South Africa\\
$^4$ IRAP, Univ. de Toulouse, CNRS, 9 avenue du Colonel Roche, 31028 Toulouse Cedex 4, France\\
$^5$ School of Mathematics \& Statistics, Newcastle University, Newcastle upon Tyne, NE1 7RU, UK\\
$^6$ ASTRON, Postbus 2, 7990 AA Dwingeloo, The Netherlands\\
$^7$ School of Physics and Astronomy, University of Southampton, Southampton SO17 1BJ, UK\\
$^8$ Chalmers University of Technology, Onsala Space Observatory, 439 92 Onsala, Sweden\\
$^9$ Dept. of Physics and Tsinghua Center for Astrophys., Tsinghua Univ., Beijing, 100084, China\\
$^{10}$ INAF-IRA, via Gobetti 101, 40129 Bologna, Italy\\
$^{11}$ Max-Planck-Institut f\"ur Astronomie, K\"onigstuhl 17, 69117 Heidelberg, Germany\\
$^{12}$ Moscow State University, Moscow, 119991, Russia\\
$^{13}$ The University of Calgary, Calgary, AB T2N 1N4, Canada
\\
E-mail: \email{rbeck@mpifr-bonn.mpg.de}
}

\abstract{
Magnetic fields are an important ingredient of the interstellar medium (ISM).
Besides their importance for star formation, they
govern the transport of cosmic rays, relevant to the launch and regulation of galactic
outflows and winds, which in turn are pivotal in shaping the structure of halo magnetic fields.
Mapping the small-scale structure of interstellar magnetic fields in many nearby galaxies is crucial to
understand the interaction between gas and magnetic fields, in particular how gas flows are affected.
Elucidation of the magnetic role in, e.g., triggering star formation, forming and stabilising spiral arms,
driving outflows, gas heating by reconnection and magnetising the intergalactic medium has the potential
to revolutionise our physical picture of the ISM and galaxy evolution in general.
Radio polarisation observations in the very nearest galaxies at high frequencies ($\ge 3$\,GHz) and
with high spatial resolution ($\le\,5''$) hold the key here.
The galaxy survey with SKA1 that we propose will also be a major step to understand the galactic dynamo,
which is important for models of galaxy evolution and for astrophysical magnetohydrodynamics in general.
Field amplification by turbulent gas motions, which is crucial for efficient dynamo action,
has been investigated so far only in simulations, while compelling evidence of turbulent fields
from observations is still lacking.

}

\FullConference{Advancing Astrophysics with the Square Kilometre Array,\\
		June 8-13, 2014\\
		Giardini Naxos, Sicily, Italy }

\newcommand{\skipthis}[1]{}

\begin{document}

\section{Introduction}

Magnetic fields are a major agent in the ISM and control the density and distribution of cosmic rays (CRs).
CRs accelerated in supernova remnants can provide the pressure to drive galactic outflows and buoyant loops
of magnetic fields via the Parker instability. Outflows from starburst galaxies in the early Universe may have
magnetised the intergalactic medium. Ultra-high-energy CRs (UHECRs) are deflected by regular
fields and scattered by turbulent fields, so that the structure and the extent of the fields in the disk and
halo are necessary parameters for propagation models.

Most of what we know about interstellar magnetic fields comes through the detection of radio waves.
The intensity of {\em synchrotron emission}\ is a measure of the number density of cosmic-ray electrons (CREs)
and of the strength of the component of the total magnetic field in the sky plane.
Linearly polarised synchrotron emission emerges from ordered fields in the plane of the sky, which can be
{\em regular (coherent) fields}, with a constant direction over scales of several kpc, or
{\em anisotropic turbulent (or tangled) fields}, reversing their direction on small scales.

High-resolution polarisation images are needed to understand the physics of interstellar magnetic fields.
Observations in the Milky Way are hard to interpret because the structures on a
range of different scales contribute to the emission and the distances are difficult to determine.
In nearby galaxies the sensitivity of present-day radio telescopes is insufficient
at high spatial resolution. Here we propose a polarisation survey of nearby galaxies with SKA1 at high
frequencies.

\section{Magnetic fields in the ISM of galaxies}

\noindent{\bf Coupling between magnetic fields and gas}\\
One of the most fundamental correlations in extragalactic astronomy is the one between the nonthermal radio
continuum and the thermal infrared (IR) luminosities of galaxies (Bell 2003). It is almost linear,
holds in all star-forming galaxies, is invariant for more than
seven orders of magnitude in luminosity (Beswick et al. 2015)
and does not evolve between redshifts of zero and about three (Murphy 2009).
Hence, radio continuum emission may be regarded as an excellent tracer of star formation at all epochs, but the
role of magnetic fields still needs to be understood.

In an optically thick (``calorimeter'') model, the linearity of the global correlation is attributed to young
massive stars, as the common source of IR and radio emission. In an ``equipartition''
model (Niklas \& Beck 1997), a tight
but slightly nonlinear correlation is obtained for a coupling between magnetic fields and total gas density.
Assuming that the radio--IR correlation is due to the amplification of turbulent magnetic fields in star-forming
regions, Schleicher \& Beck (2013) modelled the evolution of the radio--IR correlation with redshift.
Since the global correlation holds for starburst galaxies as well as for low-surface-brightness galaxies, Bell (2003)
and Lacki et al. (2010) suggested that the correlation is due to a conspiracy of several factors.

Remarkably, the radio--IR correlation also holds locally within galaxies,
which can be explained by the magnetic field--gas coupling (Tabatabaei et al. 2013a).
To date, observations in a few nearby galaxies show that
(1) the exponent of the local correlation is significantly smaller than that of the global correlation,
probably due to CRE diffusion (Berkhuijsen et al. 2013),
(2) the local correlation holds not only in star-forming
regions but also in interarm regions and outskirts with low star formation rates per surface area
(Tabatabaei et al. 2013a),
(3) the slope of the local correlation varies from region to region (Dumas et al. 2011, Basu et al. 2012), and
(4) the local correlation breaks down below a certain scale that may be interpreted as
the diffusion length of CREs (Tabatabaei et al. 2013b).
High-resolution radio continuum observations in nearby galaxies are required to assess the models.

The ratio of IR/radio intensities is a measure of current turbulent field amplification in star forming
regions (Tabatabaei et al. 2013b). It is not constant across a galaxy's disk but changes with both
the surface density of the star formation rate and the magnetic field strength
(Murphy et al. 2008, Heesen et al. 2014).
Sensitive and high-resolution SKA observations of nearby galaxies will shed light on the amplification of
turbulent magnetic fields in different environments.

The correlation between the non-thermal radio continuum and the molecular gas (traced via its CO line emission),
observed with 60\,pc resolution in the spiral galaxy M51 (Schinnerer et al. 2013), is tighter than the
correlation between radio and IR emission, but even less well understood.
Much of the molecular gas in the spiral arms of M51 is not directly associated with ongoing massive star formation,
implying that star formation cannot be the sole cause for the radio--CO relation.
Two scenarios are consistent with the M51 data:
(1) increased synchrotron emission from secondary CREs produced
in the interaction of the CRs with the dense molecular material (e.g. Murgia et al. 2005)
or (2) enhanced radio emission due to a coupling between the gas density and total magnetic field strength
(e.g. Niklas \& Beck 1997), in connection with fast diffusion of CREs along the ordered magnetic fields
of the spiral arms.
The proposed SKA1 survey of nearby galaxies with improved resolution in conjunction
with high-quality data on molecular gas disks from
the Atacama Large Millimeter/Submillimeter Array (ALMA) can provide the necessary information to test
which of the proposed origins for this correlation is indeed true. Such observations
of the magnetic field properties can be combined with exquisite knowledge of the multiple phases of the
molecular gas disk (probed via different molecular lines). Understanding the relation between magnetic
fields and molecular gas is critical, as magnetic fields are expected to play a major role in the structure
formation of the molecular ISM, and in particular the formation of giant molecular clouds that are the
cradles of massive star formation (e.g. Federrath \& Klessen 2013).\\

\noindent{\bf Magnetic fields and gas dynamics}\\
Magneto-hydrodynamic (MHD) calculations in 2-D indicate that perturbations in the gravitational potential
can excite slow and fast MHD density waves in thin galactic disks (Lou \& Fan 1998) that may explain the spiral field
patterns in galaxies like NGC6946 and M51. Fast MHD density waves are expected to be important for enhancing
MHD winds emerging from spiral arms (Lou \& Wu 2005). The role of slow MHD density waves is still unclear because they
may become unstable in thick galaxy disks and develop into Parker loops.
Gaseous spiral arms are smoother and better defined when strong magnetic fields
are included because instabilities are suppressed (Dobbs \& Price 2008).
Global MHD modelling gives evidence that galaxy evolution is modified by magnetic fields (Pakmor \& Springel 2013).
With the advent of SKA1, the interaction between density perturbations and magnetic fields can be investigated
in detail with help of the kinematic data of neutral and ionised gas.

In galaxies with massive bars the magnetic field is aligned with the gas flow and is strong enough to affect the flow.
As the gas rotates faster than the bar pattern of a galaxy, a shock occurs in the cold gas and the total magnetic
field, which is observable as a ridge of strong radio continuum
emission (Fig.~1 left). The warm diffuse gas has a larger sound speed and hence is only slightly compressed.
The ordered magnetic field is coupled to the diffuse gas. The polarisation pattern in barred galaxies
and around central bars of spiral galaxies (Fig.~2) can be used as a tracer of shearing gas flows
in the plane of the sky and hence complements kinematic data of radial velocities, but no detailed comparison can be
performed with the sensitivity of present-day telescopes.

The central regions of barred galaxies are often sites of ongoing intense star formation and strong magnetic fields
that can affect gas flows, with NGC1097 as a prominent example (Fig.~1 left).
The ordered field in the ring has a spiral pattern
and extends towards the active nucleus. The orientation of the innermost spiral field agrees with that of the
spiral dust filaments visible in optical images. Magnetic stress in the circumnuclear ring due to the strong total
magnetic fields (50--100\,$\mu$G) or fast MHD density waves can drive gas inflow at a rate of several solar masses
per year, which is sufficient to fuel the activity of the nucleus (Lou et al. 2001, Beck et al. 2005).
The infall rate is expected to be related to the field strength and structure, which can be tested with SKA1
in a sample of barred galaxies.\\

\begin{figure}[htbp]
\includegraphics[width=0.46\textwidth]{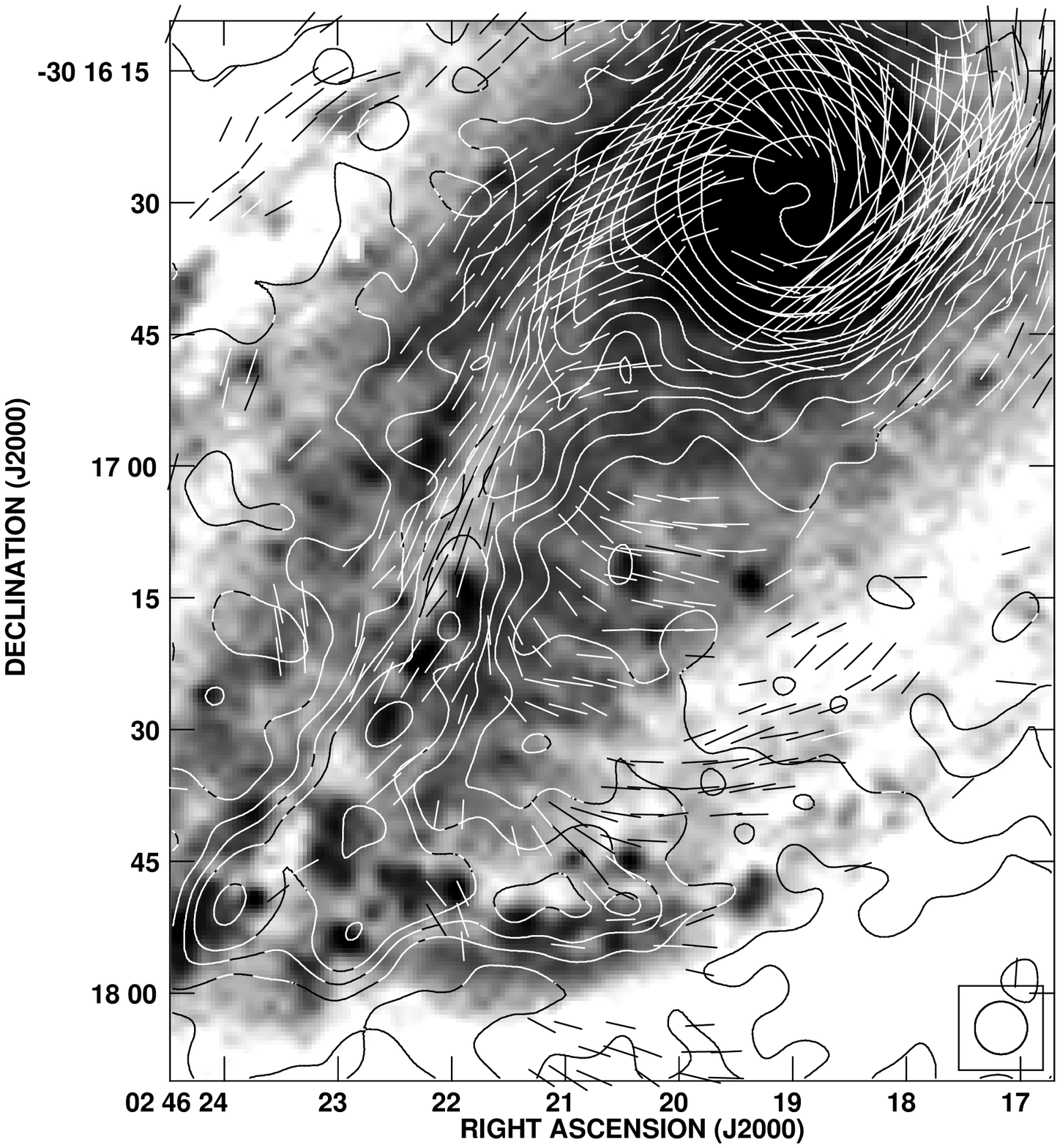}
\hspace{0.3cm}
\includegraphics[width=0.51\textwidth]{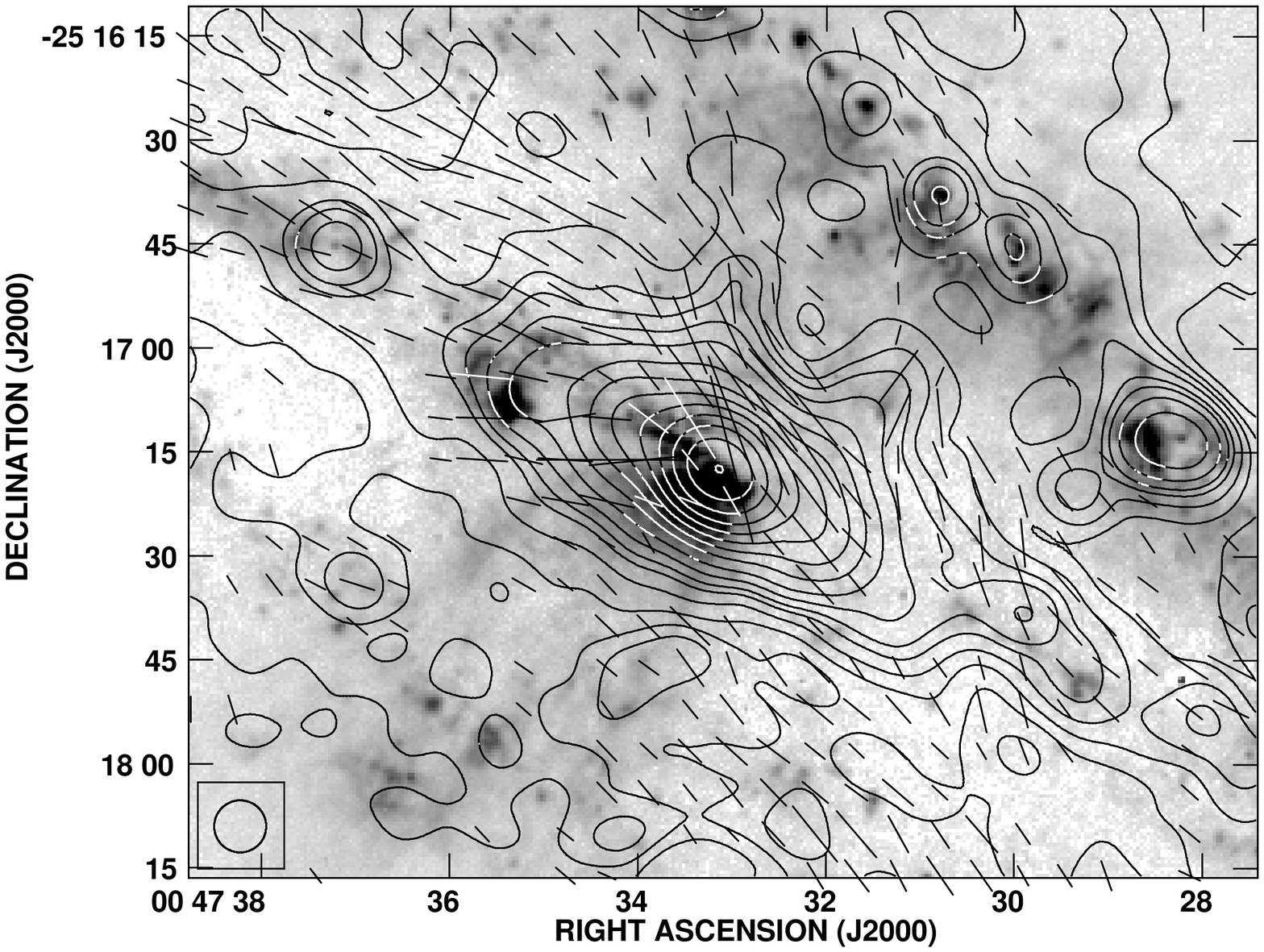}
\caption{Examples for high-resolution polarisation observations.
{\it Left:\/} Total intensity (contours) and $B$-vectors of the southern bar of NGC1097, observed at 4.86\,GHz
at $6''$ resolution with the VLA (from Beck et al. (2005)), overlaid onto an optical image.
{\it Right:\/} Total intensity (contours) and $B$-vectors of the central starburst region of NGC253, observed
at 8.46\,GHz at $7.5''$ resolution with the VLA (from Heesen et al. (2011)), overlaid onto an H$\alpha$ image.
}
\end{figure}

\noindent{\bf Ordered fields and field reversals}\\
Ordered fields, observed by polarised radio emission, are not spatially correlated with
spiral arms seen in the optical and IR (see example in Fig.~2).
They are modified by large-scale dynamical effects like galaxy rotation and density waves, as observed in M51
(Fletcher et al. 2011).
Ordered fields can be strongest in shear or compression regions, caused by gas flows (as observed in barred galaxies,
see Fig.~1 left), in regions with little
star formation by enhanced mean-field dynamo action, or in regions with moderately
strong outflows by allowing for more efficient mean-field dynamo action (see Sect.~4).
Ordered fields are generally strongest in the regions {\em between}\ the optical spiral arms,
in some cases forming {\em magnetic arms} with exceptionally
high degrees of polarisation (up to 50\%), e.g. in NGC6946 (Beck 2007), NGC2997 (Han et al. 1999),
M83 (Beck \& Wielebinski 2013) and IC342 (Krause 1993; Fig.~2).
Proposed explanations are slow MHD density waves (Lou \& Fan 1998), shear in a spiral gas flow
(Otmianowska-Mazur et al. 2002)
or suppression of large-scale dynamo action in the spiral arms
(Moss et al. 2013, Chamandy et al. 2015).

Shearing and compressing gas flows make turbulent fields anisotropic, so that the field direction reverses on the
correlation scale of turbulence (see below), while dynamo action generates uni-directional regular fields.
As polarisation ``vectors'' are ambiguous by $180^\circ$, Faraday rotation measures (RMs) are the only way
to distinguish these two types of ordered fields.

Large-scale field reversals at certain distances from the center of a galaxy, like that observed in the Milky Way
(Van Eck et al. 2011, Han et al. 2015), have not been detected in external galaxies so far, in spite of
many RM data with sufficiently high resolution. A satisfying explanation is still lacking. The reversals
may be due to propagating dynamo waves (Ferri\`ere \& Schmitt 2000), relics from the early epoch where the field pattern
was chaotic (see Sect.~4) or field distortions triggered by a major merger (Moss et al. 2014).
The search for reversals in external galaxies, and to measure their number and scale, needs RM data
with high spatial resolution of less than 100\,pc.\\

\begin{figure}[htbp]
\begin{center}
\includegraphics[width=0.7\textwidth]{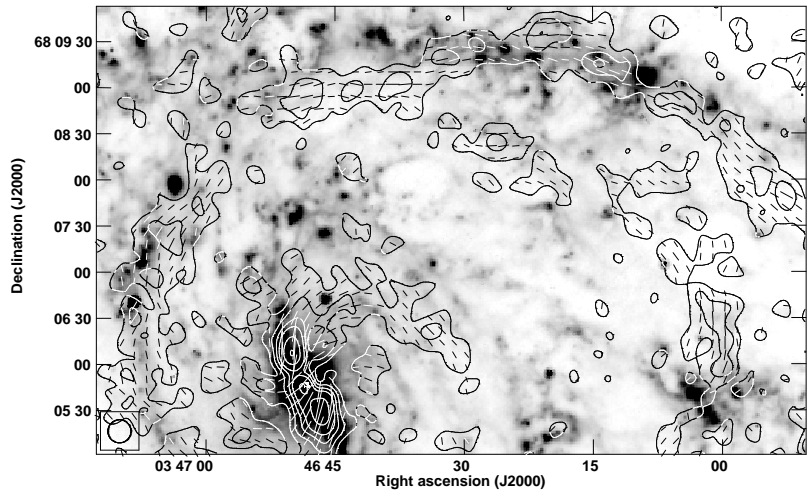}
\caption{Polarised intensity (contours) and $B$-vectors of IC342, observed at 4.86\,GHz
at $15''$ resolution with the VLA (from Beck (2015)), overlaid onto an 8$\mu$m IR image from the SPITZER telescope.
Ordered fields can be traced around the central bar, along the inner edges of the eastern and northern spiral arms, and
in the interarm region in the northwest.
}
\end{center}
\label{fig}
\end{figure}

\noindent{\bf Properties of turbulent magnetic fields}\\
Turbulent magnetic fields have their origin in the turbulent state of the ISM. The power law of
turbulence of the ionised gas in the Milky Way is of Kolmogorov type and spans to a largest scale of at least a
few pc (Chepurnov \& Lazarian 2010). To understand the origin and role of turbulent fields, they should be described using
measures derived from the theory of turbulence. Such fundamental properties are currently
unconstrained by observations, apart from a few estimates.

Structure functions of RMs in the Milky Way indicate that the correlation scale (the largest
scale of the power law of turbulence) is about 20\,pc in spiral arms and about 100\,pc in interarm regions
(Haverkorn et al. 2006). Faraday depolarisation in nearby galaxies can be explained by a turbulent
medium with a correlation scale of about 50\,pc (Beck 2007, Fletcher et al. 2011), but the spatial resolution of
present-day observations is insufficient to probe such scales directly.

The method for deriving correlation scales from the scatter in the observed polarisation angles at high
radio frequencies (where Faraday rotation angles are small) enables measurement of the degree of anisotropy
in magnetic field fluctuations (Houde et al. 2013). The application of this technique is currently limited to a few
bright patches of polarisation seen at the highest possible resolution in M51, where the
correlation lengths parallel and perpendicular to the local ordered field are about 100\,pc and 50\,pc,
which may lead to anisotropic propagation of CREs (see below).


Polarisation observations with SKA with high spatial resolution (i.e. 1--100\,pc) and at high frequencies
(to avoid Faraday depolarisation) will allow the measurement and mapping of standard quantities,
such as the correlation
scale and degree of anisotropy and the power spectrum of the turbulent field.
It will then be possible to address the origin and role
of the turbulent field, for example, in the accumulation of gas clouds from the diffuse ISM, and to determine
the sonic type of turbulence from polarisation gradients (Burkhart et al. 2012).\\

\noindent{\bf Helical fields}\\
Magnetic helicity is a promising concept in the description of astrophysical magnetic fields.
A significant fraction of magnetic fields in the ISM may be helical, but firm observational evidence is still lacking.
The reversing RM sign across the outflow cone of NGC253 indicates a large-scale helical
field (Heesen et al. 2011). Polarised emission and RMs in the northern radio arm of the galaxy
IC342 reveal a helically twisted field (Fig.~2). Simulations of galactic dynamos (see Sect.~4) predict the
formation of bi-helical magnetic fields of opposite helicity on large and small scales (Brandenburg 2015).
Such fields leave an observational signature in the spectral energy distribution of the polarised signal.
A suitable sign of helicity can compensate Faraday depolarisation.

High-resolution polarimetric observations with SKA1 will be very useful to probe the topology of galactic
magnetic fields and place constraints on their helicity (Horellou \& Fletcher 2014).
According to dynamo theory (Sect.~4), the toroidal field should have the same direction above
and below the midplane,
while the magnetic helicities would change sign about the equatorial plane. For edge-on galaxies, enhanced
polarisation can be expected in two opposite quadrants (Brandenburg \& Stepanov 2014).
Large-scale helical fields in galaxies may also be
revealed by investigating the correlation between polarisation angles and RMs
(Oppermann et al. 2011).\\

\noindent{\bf Magnetic effects on the propagation of cosmic-ray electrons}\\
The degree of order of the magnetic field plays an important role in the propagation of CREs, both in models of
the streaming instability (Kulsrud 2005) and in diffusion models
(Chuvilgin \& Ptuskin 1993). A higher field ordering helps CREs to propagate over larger distances along the field,
resulting in a larger propagation length. This is in agreement with the observations in a sample
of nearby galaxies, where the smallest scale of the synchrotron--IR cross-correlation increases with field
order (Tabatabaei et al. 2013b). With future high-resolution SKA1 data, cross-correlation for different
regions of a galaxy will allow us to measure the CRE propagation length as a function of environmental conditions,
in particular of the magnetic fields.\\

\noindent{\bf Irregular and dwarf galaxies}\\
The Large and the Small Magellanic Clouds are the closest irregular galaxies at distances of 50\,kpc and 60\,kpc,
respectively, interacting with the Milky Way and each other.
A study based on RM grids and thus on line-of-sight magnetic field components suggests that both
galaxies host large-scale coherent fields, indicating that a large-scale dynamo also works under the
less favourable conditions of slow ordered rotation (Mao et al. 2008).
High spatial resolution (about 1\,pc!), high-frequency SKA1 observations as proposed here can overcome
beam and Faraday depolarisation, allowing one to detect diffuse polarised emission from the main body of the
Magellanic Clouds.
We can test with unprecedented resolution how well the polarised filaments and their ordered
field align with various tidal features in the Magellanic
system (the so called Pan-Magellanic magnetic field hypothesis). The proposed observations of the Magellanic
Clouds will also allow us to construct the structure function of the complex polarised emission
(Sun et al. 2014) to understand the turbulence cascade and dissipation.

With their low rotational speed dwarf galaxies are even less favourable places for
the amplification of magnetic fields. Still, several
low-mass galaxies (besides the Magellanic Clouds) show a significant
large-scale ordered field in some regions, like e.g.\ IC10 (Chy{\.z}y et al. 2003).
Models of a CR-driven dynamo (Siejkowski et al. 2014) and MHD simulations (Dubois \& Teyssier 2010)
imply that dwarf galaxies can amplify fields and transport them away in galactic outflows and winds.
Still, many details are unexplored, like the effects of magnetic fields early
in the formation of the galactic outflow and their role in collimating
the outflowing gas.

Due to their low potential well, low-mass galaxies are prone to outflows and
galactic winds driven by star formation.
The magnetised nature of these large-scale outflows
(Chy{\.z}y et al. 2000, Kepley et al. 2010) is especially noteworthy, since dwarf irregular
galaxies are the best local proxies for
high-redshift galaxies and even proto-galaxies and a prime candidate for the source of the
magnetisation of the intergalactic medium (Bertone et al. 2006).
An important observation is that all dwarf galaxies
with detected ordered magnetic fields are star-bursting and/or subject to
gravitational interaction or gas infall. This may be an observational
selection effect but may also hint at the importance of enhanced turbulence in these galaxies.

SKA1 observations of a sample of local dwarf galaxies exploring the effects of burst age,
burst strength, mass and interaction properties will shed light on the
mechanisms governing the amplification and dispersion of magnetic fields in
such galaxies. These
observations will also provide critical data for the effects
of magnetic fields during galaxy formation and early evolution.\\

\noindent{\bf Early-type and elliptical galaxies}\\
Spiral galaxies of type Sa and S0 as well as elliptical galaxies without an
active nucleus have very little star formation and hence hardly
produce CREs that could emit synchrotron emission. Large-scale
regular magnetic fields may exist in differentially rotating early-type galaxies
without star formation, because turbulence in the ISM can also be generated
by the magneto-rotational instability (MRI). Elliptical galaxies may host turbulent
fields generated by the small-scale dynamo. Detection with SKA1 via RM grids
needs a sufficiently large number of background sources and hence deep observations
of galaxies with large angular sizes.

\section{Magnetic fields in galaxy halos}

\noindent{\bf Outflows}\\
Galaxies are not isolated entities, but they can exhibit both accretion and outflow of the gas that fuels star
formation. Galactic wind or outflow models can reproduce the available multi-frequency observations only if
CRs and magnetic fields help to activate the wind at their base in galactic bulges (Everett et al. 2008).
Although CRs were neglected in most earlier galactic wind models, their energy density is in fact comparable to the
respective values of energy densities for thermal gas, magnetic fields and turbulent motions. This energy is
concentrated in protons and heavy nuclei, while CREs contribute only $\sim 1\%$.
Compared with thermal gas, CR protons do not cool efficiently while CREs cool dominantly through inverse
Compton scattering (ICS) and synchrotron emission. The combined pressure gradients of gas, CRs and MHD
waves are able to drive a galactic wind even in cases of moderate star formation, like in our own Galaxy
(Breitschwerdt et al. 1991). A necessary condition is that there is a dynamical coupling between
the plasma and the CRs.

If the population of CREs consists of secondary particles of hadronic origin, the $\gamma$--ray emission
from galaxy winds or outflows is dominated by hadronic emission mechanisms with a typical pion bump originating
from $\pi^0 \to \gamma \gamma$ decay that dominates the GeV energy range, while secondary electrons
produce GeV--TeV ICS emission and relativistic bremsstrahlung. In a leptonic-origin model for CREs,
ICS emission from primary CREs dominates in the GeV--TeV range.

Therefore, the radio emission must retain the imprints of either hadronic (secondary CREs) or leptonic
(primary CREs) produced synchrotron emission with specific spectral shapes and radio polarisation features.
The correlation between radio (continuum and polarisation) and $\gamma$--ray (GeV--TeV) emission could provide
valuable information on the CR origin and propagation and on the magnetic field that confines these particles
in the galaxy winds and halos produced by outflows (see Sect.~5).
In addition, polarisation from primary CREs is expected to be higher, because secondary CREs are produced
in a more isotropic environment.

The distribution of the multi-temperature plasma and of the non-thermal (relativistic) plasma in the winds
provides a sea of particles available to produce ICS of the CMB photons pervading all of the
Universe and therefore an associated Sunyaev-Zeldovich effect (SZE) with specific spatial and spectral
characteristics that are related to the complex plasma distribution in the winds (Colafrancesco et al. 2003).
The combination of radio synchrotron emission from CREs (detected with SKA) and the SZE in the same
wind/outflow region as produced by the same CREs that up-scatter the CMB photons (detectable with mm and
submm telescopes) can be used to obtain a direct estimate of the overall magnetic field
energy density $U_\mathrm{B}$ in the wind or outflow region by measuring the ratio of integrated flux densities
$F_\mathrm{sync}/F_\mathrm{ICS} \propto U_\mathrm{B}/U_\mathrm{CMB}$.
While edge-on galaxies are optimal targets (e.g. NGC253), galaxies with disks seen at intermediate
inclination angles are also suitable.\\

\noindent{\bf Outflow velocities}\\
CRs (the most pervasive of
the ISM components) can easily leave the disk either by diffusion and streaming along magnetic
field lines or by convective transport in a galactic outflow or wind. Indeed,
many edge-on galaxies do clearly exhibit signatures of extra-planar emission
and hence the presence of radio halos surrounding them (Krause 2014).
Gravitational interaction with a companion galaxy may affect the outflow (Mora \& Krause 2013).

Vertical profiles of the radio continuum emission can be well described by a two-component
exponential distribution (e.g. Dumke \& Krause 1998, Mora \& Krause 2013), where the
two scale heights describe the thin and the thick disks; the latter can be attributed to the radio halo.
As the energy densities of the magnetic field and the stellar background radiation are higher in the inner parts
of a galaxy, the CRE lifetimes and hence the scale heights should have a minimum at small galacto-centric radii.
This is indeed observed in a few galaxies (Krause 2014). The most striking example is NGC~253,
which displays a dumbbell-shaped halo. The linear dependence between scale height and CRE lifetime in its
northeastern halo is a signature of a wind with a cosmic-ray bulk speed of $300\pm30~\rm km\, s^{-1}$
(Heesen et al. 2009a), which is similar to the escape velocity in this galaxy. Measuring the velocity of outflows
and its variation with height above the plane from the vertical profiles of radio emission
needs high-resolution data of a sample of edge-on galaxies.

On a spatial scale of a few tens of parsec, accessible to SKA1,
we expect that the disk--halo interface will be resolved into many individual features, among them non-thermal
super-bubbles, which are crucial to drive a galactic outflow by supernova feed-back.\\

\noindent{\bf X-shaped fields}\\
The large-scale magnetic field in nearby spiral galaxies is not only confined to the galactic disk.
Polarisation observations of spiral galaxies seen edge-on show a projected plane-parallel magnetic field
in the disk and an X-shaped field structure of similar field strength in the halo
(Heesen et al. 2009b, Soida et al. 2011, Krause 2014).
According to mean-field $\alpha-\Omega$ dynamo models (see Sect.~4), the regular disk
field is accompanied by a poloidal magnetic field component reaching into the halo, which,
however, is far too weak to explain the observed X-shaped field.
Some similarity to the observed field patterns is achieved if the effects of a galactic wind are included
(Gressel et al. 2008, Hanasz et al. 2009).
The X-shaped field may also be a superposition of plane-parallel and vertical fields composed
of field loops stretched by outflows. Frequent field reversals between adjacent field loops may lead to
turbulent reconnection and heating of the halo gas.

Faraday RMs are the adequate tool to test observationally whether the halo field is regular
(large $|RM|$) or composed of loops (small $|RM|$), but reliable measurements in galaxy halos are still missing.
Making use of the high resolution and sensitivity of SKA1 will settle this issue.

\section{Tests of magnetic field models with SKA1 observations}

\noindent{\bf Mean-field dynamo models of galactic magnetic fields}\\
The best developed paradigm to amplify and sustain magnetic fields in the ISM
of galaxies is the dynamo (Beck et al. 1996, Brandenburg 2015). A
turbulent (small-scale) dynamo in protogalaxies can amplify weak seed fields
to several $\mu$G strength (corresponding to the energy level of turbulence) within
less than $10^8$\,yr (Schleicher et al. 2010, Beck et al. 2012).
After formation of a gas disk, the small-scale dynamo
can continuously supply turbulent fields to the ISM (Moss et al. 2012).
To explain the generation
of large-scale fields in galaxies, the theory of mean-field $\alpha-\Omega$ dynamo has been
developed. It is based on differential rotation ($\Omega$--effect) and helical turbulence
($\alpha$--effect), presumably driven by supernova explosions (Ferri\`ere \& Schmitt 2000, Gent et al. 2013),
and predicts that within a few $10^9$\,yr turbulent fields of a few $\mu$G strength are
organised into large-scale patterns of regular fields (Arshakian et al. 2009).
Field reversals from the early epochs can survive until today (Shukurov 2005, Moss et al. 2012)
and may explain the large-scale field reversal(s) detected in the Milky Way (Van Eck et al. 2011).

The mean-field $\alpha-\Omega$ dynamo generates
large-scale magnetic helicity with a non-zero mean in each hemisphere.
As the total magnetic helicity is a conserved quantity, the dynamo is quenched by
the small-scale fields with opposite helicity unless these are
removed from the system (Shukurov et al. 2006). Galactic outflows (see Sect.~3) are essential
for efficient $\alpha-\Omega$ dynamo action.




Mean-field $\alpha-\Omega$ dynamo models predict a regular field with a simple axisymmetric pattern, whereas
regular fields in real galaxies have complicated patterns that may be described by several Fourier modes
(Fletcher 2011). High-resolution observations with SKA1 will allow us to measure the power spectrum
(or structure function) of regular fields, from the
turbulence scale up to the largest scale the dynamo was able to generate within the galaxy's lifetime.
Signature of a fully developed regular field is a peak in the power spectrum at the scale of the whole galaxy
(Moss et al. 2012, Gent et al. 2013). Field reversals, density waves, bars and mergers leave specific imprints
in the power spectrum.\\

\noindent{\bf Connection between magnetic fields, galaxy properties and dynamo action}\\
Mean-field dynamo models depend on observable properties of galaxies. The rotation curve
determines the strength of the shear that generates an azimuthal field from the radial field, while the
star formation rate governs the frequency of supernova explosions driving the $\alpha$--effect
by which a radial field is generated from the azimuthal field.
Gas density and turbulent velocity determine the energy density of turbulence which provides
an upper limit on the energy density of the magnetic field.
The timescale for large-scale field ordering
can be longer than the galaxy age (Arshakian et al. 2009). The stage of development of a
field pattern also carries information about the history of a galaxy, e.g. the
time and strength of the last major merger (Moss et al. 2014).

Determining the
relationships between these well-defined galactic properties and the strength and structure of dynamo-generated
magnetic fields will provide a significant step forward in both our understanding of cosmic magnetism and the
application of this knowledge to problems in wider astrophysics and cosmology. For example, knowing how
turbulent and regular field strengths scale with star formation rate and gas density will immediately
allow the inclusion of magnetic fields in analytic and semi-analytic models of galaxy evolution.

Identifying the precise form of the connection between magnetic and other galactic properties is not straightforward
because the parameters can combine to produce non-trivial
scalings. This difficulty will fortuitously provide another significant result though. Different physical
mechanisms have been proposed to provide the non-linearity which is required to saturate the field growth,
or allow it to reach a quasi steady state.
Examples of the saturation mechanisms are: the back-reaction of the Lorentz force on the turbulent motions
when it becomes comparable to the driving forces of turbulence;
outflows of hot gas
carrying small-scale magnetic fields out of the disk in order to preserve the balance of magnetic helicity.
These non-linear saturation mechanisms depend differently on galactic
parameters such as star formation rate and gas density (Van Eck et al. 2014).
Thus, determining the fundamental scaling relations between
the observed magnetic fields and the other galactic parameters will also allow us to identify the saturation
mechanism for the dynamo.

Understanding the non-linear behaviour of galactic dynamos will be of major importance for the wider
field of cosmic magnetism, because virtually all astronomical objects are magnetised
by dynamo action.
Galaxies are special though: They are the only objects where the dynamo-active region is transparent and
can be directly observed.

The main problem in carrying out this work now is the lack of suitable magnetic field observations for a reasonable
number of galaxies. What is required is a set of data obtained with the same instrument,
with the same spatial resolution in each galaxy, the same sensitivity and analysed using the same methods to obtain
a reliable catalogue of comparable magnetic field strengths, rotational and vertical field symmetries,
and pitch angles.\\

\noindent{\bf Origin of small-scale magnetic fields in galaxies}\\
Aside from the regular component, galactic magnetic fields have a small-scale, fluctuating component.
The turbulent mirror-asymmetric motions perform twisting and folding of magnetic loops
stretched by differential rotation. The same turbulent motions tangle the
regular magnetic fields and produce small-scale magnetic fields from the large-scale ones.

Dynamo theory suggests another process: the turbulent dynamo, where stretching, twisting and folding
of magnetic loops are produced by random motions. Hence, the total magnetic field $B_\mathrm{tot}$ can be
presented as a sum of the regular field $B_\mathrm{reg}$, the component $B_\mathrm{tan}$ produced by
tangling of $B_\mathrm{reg}$ and a component $B_\mathrm{tur}$ generated by the turbulent dynamo.
An equipartition argument suggests that $B_\mathrm{tan}$ and $B_\mathrm{tur}$ should be
comparable. Sunspot statistics demonstrated that $B_\mathrm{tur}$ is negligible with respect to
$B_\mathrm{tan}$ on the solar surface (Stenflo 2012).

This issue is of fundamental
importance for cosmic MHD: Without a turbulent dynamo the seed field in proto-galaxies would be much weaker
and the generation of large-scale fields would take much longer. Hence, it is important to search for
turbulent fields in galaxies, where the regime of dynamo action is fully accessible.
The magnetic field component $B_\mathrm{tur}$ can be separated from $B_\mathrm{tan}$ by high-resolution
observations with SKA1. Firstly, the field $B_\mathrm{tan}$ can be generated only in regions where $B_\mathrm{reg}$
exists. If somewhere $B_\mathrm{reg}$ vanishes, $B_\mathrm{tan}$ decays rapidly and is expected to be small.
Observational tests require a resolution of the order of the turbulent scale,
10--100\,pc, in order to resolve the vicinity of the point at which $B_\mathrm{reg}$ vanishes.
Secondly, the spatial spectrum of $B_\mathrm{tan}$ is expected to extend
from the scale of the regular magnetic field, at several kpc, down to smaller scales.
The spectral slope may also be modified, but no model predictions are available.
In contrast, the spectrum of $B_\mathrm{tur}$ extends from the largest scale
of interstellar turbulence, at about 100\,pc, down to about 10\,pc and possibly lower.
If both components contribute to the resulting spectrum of small-scale magnetic fields,
a spectral distinction is possible by high-resolution observations.

\section{Synergy with observations in other spectral ranges}


\noindent{\bf Dust}\\
The infrared space observatories SPITZER and HERSCHEL have mapped the emission of the
dust continuum and also provided exquisite maps of the major ISM cooling lines for several tens of nearby
galaxies (e.g. Kennicutt et al. 2011).
In addition, the advent
of new bolometer cameras for existing millimeter single dishes such as APEX and JCMT, as well as planned ones such
as CCAT (or LMT), will allow for high angular resolution imaging of the cold dust component at wavelengths
long-ward of about $200\,\mu$m with the possibility of polarisation measurements.\\

\noindent{\bf Molecular gas}\\
Millimeter and submm observatories (JCMT, APEX, ALMA, PLANCK) have the capacity to map both the flat-spectrum,
high-frequency radio emitting regions and the extended halos around normal and starburst galaxies.
Future observatories with improved angular resolution
and sensitivity (e.g., the ground-based SPT, ACT telescopes and especially the large Millimetron space mission with
a sensitivity matching that of SKA1) can provide a great deal of synergy with the SKA1 and SKA2 studies of nearby
and moderately distant galaxies.

ALMA is going to revolutionise
our understanding of the molecular ISM in local galaxies via high angular resolution observations of the
multiple phases present in the molecular gas and the gas kinematics. During its early science phase ALMA has
already begun to map the distribution and kinematics of the bulk of the molecular gas in the central regions
of a few nearby galaxies.
With the full array operational it will be possible to extend such studies not only to more targets probing
a wide range in galactic disk properties (quiescent -- star forming; flocculent -- barred/spiral arms;
gas-rich -- gas-poor) but also to probe the denser molecular gas that is more closely linked to the actual
formation of stars. In addition, polarisation measurements of continuum emission arising from dust
and of line emission will become possible (Paladino et al. 2015).\\

\noindent{\bf Neutral gas}\\
SKA1 observations of the neutral ISM in nearby galaxies (de Blok et al. 2015)
will provide exquisite probes of the properties
of cold gas on both small and large scales. In addition to the
morphology of the neutral ISM, HI observations also provide highly
detailed information about the kinematics. HI line widths can be used
to probe small-scale gas motions and draw connections to the turbulent
magnetic field structure. Neutral hydrogen can also be traced out to
tremendous distances from sites of star formation, yielding
large-scale kinematic information including rotation, shear and
radial motions. These will provide crucial insight into the influence
of dynamics on the local orientation and degree of ordering of the
magnetic field in arms, interarm regions and bars, and on dynamo action.
HI is also a highly useful probe of the connection between galaxies and their
environment, including tidal interactions, mergers and gas inflow and
outflow. The excellent column density sensitivity afforded by SKA1
will illuminate the connection between large-scale magnetic field
properties and gas conditions, local dynamics and environmental effects.\\

\noindent{\bf Ionised gas}\\
The origin of the warm (or diffuse) component of the ISM in spiral galaxies can be
traced back to star formation regions, even though substantial parts of this gas are found at
large distances from star forming sites. This can be explained by the porosity of the ISM and by
large-scale flows caused by the energy input from young stellar populations. Both these explanations
depend on the contribution from magnetic fields and CRs to the total pressure of the ISM, since
these constituents may contribute about 2/3 of the energy density. To explain the observed temperature
structure of the ionised gas, an additional, so far unidentified, heating source is required.

CRs coupled to the ionised gas via magnetic fields could cause this additional heating (Wiener et al. 2013).
The new high-sensitivity tunable filter instruments and wide-field integral-field
spectrographs (e.g. MUSE) at 4--8\,m-sized telescopes will provide more sensitive observations
of diagnostic lines and the kinematics for the halo faint gas. These optical/NIR data sets
together with SKA observations at high angular resolution will address the open questions.\\

\noindent{\bf X-rays and $\gamma$-rays}\\
The relativistic electrons that produce radio synchrotron emission
also produce X-ray/gamma-ray emission through bremsstrahlung and ICS.
In principle, if the high-energy emission could be used to determine
the relativistic-electron density, then the magnetic field strength could be
directly inferred from the radio synchrotron data, without having to appeal
to the equipartition assumption.
In practice, though, X-ray/gamma-ray telescopes lack the required resolution and sensitivity
and the high-energy X-ray emission from relativistic electrons is hard to separate from
other emission components ($\pi^0$ decay, thermal emission and point sources).
Halos of edge-on galaxies are preferable targets because their X-ray emission is mostly thermal.
On the other hand, it should be possible to infer the un-normalized
relativistic-electron spectrum from the measured radio spectral index
and use it as a constraint to interpret the high-energy data.
This synergy between both spectral ranges
would work best outside the 100\,MeV--30\,GeV range, which is
largely dominated by $\pi^0$ decay.

On the low-energy side, the most promising telescope could be the Nuclear
Spectroscopic Telescope Array (NuSTAR), which was launched in June of 2012,
covers the energy band 3--79\,keV and has an angular resolution of $18''$.
In this energy band, inverse Compton dominates the diffuse interstellar
emission from spiral galaxies, but thermal emission is probably dominant
both in elliptical galaxies, where the gas temperature is higher ($\approx 10^7$\,K)
while the density of relativistic electrons is generally very low,
and in galaxy clusters, where the gas temperature can be even higher ($\approx 10^8$\,K).
The main difficulty with the NuSTAR energy range is contamination by point
sources -- a difficulty which could possibly be alleviated in galaxies
with high star formation rates. In this perspective, the starburst galaxy NGC253 will be one of our
primary targets.

On the high-energy side, the most appropriate telescope would certainly be
the Cherenkov Telescope Array (CTA), which is expected to become operational
around 2020. It covers the energy band from a few 10\,GeV to about 100\,TeV and has
an angular resolution $\gtrsim 1'$. A big advantage of the CTA energy range
is that the measured emission is most likely dominated by ICS,
with little contamination from point sources up to at least 1\,TeV.
Note, however, that the background photons that can be up-scattered to TeV or sub-TeV energies
with GeV electrons (those responsible for the radio synchrotron emission measured by SKA)
need to be higher-energy photons.

The recent upgrade of the High Energy Stereoscopic System (HESS) telescope and its enhanced sensitivity to lower
energies will provide a useful exploration of a few nearby galaxies in the sub-TeV range that
can create interesting synergies with the SKA precursors.

\section{Project description}

\noindent{\bf Sample selection}\\
With SKA1, we wish to observe a large-enough sample of star-forming galaxies that are sufficiently nearby
to achieve a high spatial resolution. A distance limit of 20\,Mpc and declinations of $\le15^\circ$ define
our preliminary sample:\\
-- Spiral galaxies (moderately inclined): M33\footnote{High declination but included due to its
proximity.}, NGC300, 628, 1566, 1808, 2997, Circinus\\
-- Edge-on galaxies: M104, NGC55, 253, 3628, 4666, 4945\\
-- Barred galaxies: M83, NGC1097, 1313, 1365, 1502, 1672, 2442\\
-- Irregular galaxies: LMC, SMC\\
-- Dwarf galaxies: NGC1140, 1705, 5253, IC4662\\
-- Non-active elliptical galaxies: NGC1404, 4697.\\

\begin{figure}[htbp]
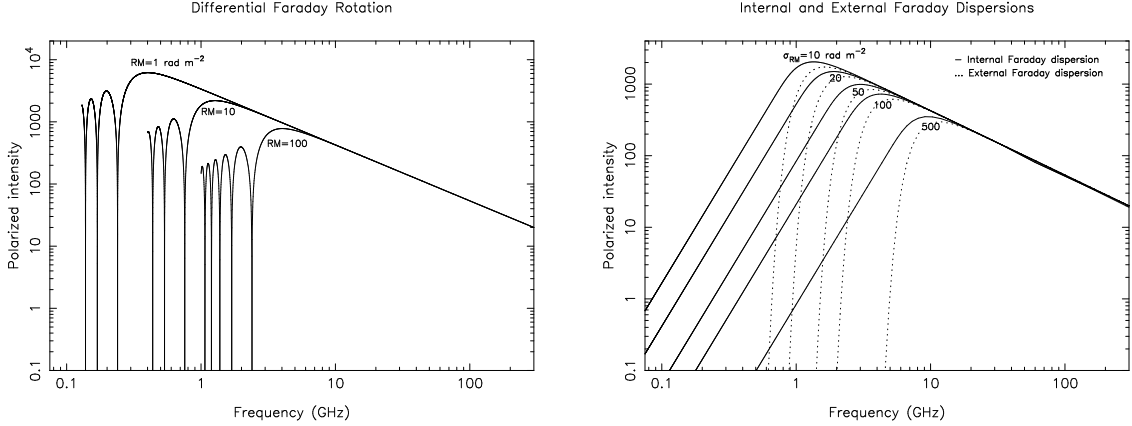

\includegraphics[width=0.37\textwidth,angle=270]{diff_far_rot.ps}
\hspace{0.7cm}
\includegraphics[width=0.37\textwidth,angle=270]{far_disp.ps}
\caption{Spectra of polarized intensity (in arbitrary units) for a synchrotron source with spectral
index $\alpha=-0.9$ (visible as a straight line at high frequencies), affected by different depolarisation
effects. {\it Left:\/} depolarisation by differential Faraday rotation at different levels of
Faraday rotation measure $|RM|$; {\it right:\/} depolarisation by internal (solid line) and external
(dashed line) Faraday dispersion at different levels of RM dispersion $\sigma_\mathrm{RM}$.
The distributions of CREs, thermal electrons, regular magnetic field and turbulent magnetic field are
assumed to be symmetric. The regular magnetic field is assumed to have a constant direction throughout the source
(from Arshakian \& Beck 2011).
}
\label{fig}
\end{figure}

\noindent{\bf Frequency selection}\\
We are aiming at:\\
-- Highest angular resolution achievable with sufficiently large signal-to-noise ratios, to detect small-scale
field structures and field reversals,\\
-- maximum polarised intensity, which requires a low level of Faraday depolarisation and hence high frequencies,\\
-- high-precision measurements of intrinsic polarisation angles with help of RM Synthesis, which requires a moderate Faraday rotation angle
and hence high frequencies (because the error in intrinsic angle increases with the Faraday rotation angle).

{\bf SKA1-MID Band~4 (2.8--5.18\,GHz)}\ fulfils our requirements:
$|RM|=50-200$\,rad/m$^2$ and RM dispersions of $\sigma_\mathrm{RM}=30-200$\,rad/m$^2$ are expected in the
disks, spiral arms and inner halos of galaxies (Arshakian \& Beck 2011). The maximum polarised intensity is expected
in the range 3--7\,GHz (Fig.~3) and a maximum rotation angle of about $\pm60^\circ$ at 4\,GHz.
We will be able to measure regions with mixed emission and Faraday rotation with $|RM|\le500$\,rad/m$^2$
(corresponding to extended structures in Faraday spectra of up to about 1000\,rad/m$^2$ width), allowing us
to detect emission regions with strong regular magnetic fields and high electron density and to
determine the intrinsic polarisation angles.

SKA1-MID Band~3 (1.65--3.05\,GHz) is less suited because the emission from galaxies with strong magnetic fields
is affected by Faraday depolarisation. SKA1-MID Band~5 (4.6--13.8\,GHz) is best suited to detect maximum polarisation
in regions with strong magnetic fields, like the central regions and star-forming complexes of galaxies, but less
suited for the general ISM and halos where the synchrotron emission is fainter.\\

\noindent{\bf Angular resolution}\\
With about $5''$ resolution we can resolve spatial scales of about 1\,pc in the
LMC/SMC, 20\,pc in M33, 100\,pc in M83, NGC55 and NGC253, sufficient to reach our goals, and a few 100\,pc in the
more distant galaxies of our sample, which is still sufficient to reach most of our goals.\\

\noindent{\bf Sensitivity}\\
With the planned System Equivalent Flux Density (SEFD) of 2.3\,Jy and the bandwidth of 2.38\,GHz of SKA1-MID Band~4
(Dewdney et al. 2013), we expect an rms noise of 0.2\,$\mu$Jy/beam
within a 12\,h integration time. This is sufficient to detect emission from a magnetic field of about 9\,$\mu$G with
$5''$ resolution, assuming an extent along the line of sight of 100\,pc, a synchrotron spectral index of $-0.8$
and energy equipartition between the energy densities of magnetic fields and CRs, or 16\,$\mu$G with 10\,pc
extent. As we aim to measure extended emission with highest sensitivity, excellent $uv$ coverage within
$\le$5\,km baselines is requested.\\

\noindent{\bf Science with a 50\% early-phase SKA1}\\
With 50\% sensitivity for SKA1-MID baselines of $\le$5\,km, an early science programme on the nearest galaxies
(LMC, SMC, M33, M83, NGC55, NGC253)
is of high scientific value, observing with a somewhat larger beam ($7''-10'')$.\\

\noindent{\bf Perspectives for SKA2}\\
The $\simeq$10x increased sensitivity of SKA2-MID compared to SKA1-MID will allow us to increase the angular resolution
by a factor of about three or to increase the sample of galaxies to about three times larger distances.
We will also consider to increase our frequency coverage towards higher frequencies, to get better access to
galactic regions with strong magnetic fields. The decision will be based on the results from SKA1.\\

\noindent{\bf Synergies with other SKA1 projects}\\
The resolution in Faraday space (given by the half-power width of the RM Spread Function) for SKA1-MID Band~4 is
about 400\,rad/m$^2$, so that only the average Faraday depth can be measured and multiple components cannot be
resolved. 3-D Faraday tomography should be done at lower frequencies (Heald et al. 2015) and perfectly
complements this project. SKA1 observations of HI emission in nearby galaxies (de Blok et al. 2015)
and in particular in the Magellanic Clouds (McClure-Griffith et al. 2015) will provide data on the properties
and dynamics of neutral gas.
The Milky Way's magnetic field will be studied with help of the RM grid and Faraday tomography
(Haverkorn et al. 2015).
The Zeeman effect will allow us to measure magnetic field strengths in star-forming regions of nearby
external galaxies (Robishaw et al. 2015).
We will get access to evolving magnetic fields in distant galaxies by deep-field observations (Taylor et al. 2015)
and by RMs against background sources (Gaensler et al. 2015).



\end{document}